# High-Energy-First (HEF) Heuristic for Energy-Efficient Target Coverage Problem

Manju[1] and Arun K Pujari[2]

[1]Department of Computer Science, Sambalpur University Institute of Information Technology, Odisha, India
`manju.nunia@gmail.com`
[2]Vice-chancellor, Sambalpur University, Odisha, India
`arun.k.pujari@gmail.com`

**ABSTRACT**

*Target coverage problem in wireless sensor networks is concerned with maximizing the lifetime of the network while continuously monitoring a set of targets. A sensor covers targets which are within the sensing range. For a set of sensors and a set of targets, the sensor-target coverage relationship is assumed to be known. A sensor cover is a set of sensors that covers all the targets. The target coverage problem is to determine a set of sensor covers with maximum aggregated lifetime while constraining the life of each sensor by its initial battery life. The problem is proved to be NP-complete and heuristic algorithms to solve this problem are proposed. In the present study, we give a unified interpretation of earlier algorithms and propose a new and efficient algorithm. We show that all known algorithms are based on a common reasoning though they seem to be derived from different algorithmic paradigms. We also show that though some algorithms guarantee bound on the quality of the solution, this bound is not meaningful and not practical too. Our interpretation provides a better insight to the solution techniques. We propose a new greedy heuristic which prioritizes sensors on residual battery life. We show empirically that the proposed algorithm outperforms all other heuristics in terms of quality of solution. Our experimental study over a large set of randomly generated problem instances also reveals that a very naïve greedy approach yields solutions which is reasonably (appx. 10%) close to the actual optimal solutions.*

## 1. INTRODUCTION

In wireless sensor networks it is often required to deploy a large number of sensors in a random fashion to monitor a set of targets. The sensors have limited battery life and it is necessary to efficiently utilize the energy while monitoring all targets for maximum duration. With the available technology sensors are battery powered and that is rarely replaced or renewed. Therefore, energy conservation is a critical issue which affects the application lifetime. As large-scale sensor networks may be overly deployed, a subset of sensors can be active at any given time to monitor all targets. Such over-deployment can be exploited to obtain a longer lifetime of battery-powered sensor network. Energy efficient target coverage problem deals with scheduling the sensor's active/sleep durations such that the total lifetime of the network is maximized while all the targets are continuously monitored. The battery life of individual sensors is the constraint that prevents the network to have arbitrarily large lifetime. The problem is shown to be NP-complete [7] and heuristics or approximate schemes are proposed in literature [5, 7, 20]. We propose a new heuristic and show that our method outperforms other algorithms and yields heuristic solutions close to the exact optimal solutions.

The major contributions of the paper are the following. We devise a new heuristic based on residual battery life of individual sensors and show that our algorithm gives closer to actual solution than that of any other methods. Experiments are carried out for a large set of randomly generated problem instances and we report a comparative analysis of our algorithm with major known algorithms.

Target Coverage Problem is introduced in section 2. Background on the earlier algorithms to solve this problem is given in section 3. Motivation behind proposed research is given in section 4. In section 5, we describe the proposed High-Energy-First heuristic. Section 6 deals with the experimental study for performance analyses. Section 7 outlines the future directions and concludes.

## 2. TARGET COVERAGE PROBLEM

Let $s_1, s_2, \ldots,$ and $s_n$ are randomly deployed $n$ sensors and $t_1, t_2, \ldots,$ and $t_m$ be m targets. Every sensor $s_i$ has a battery life of $b_i$. A sensor $s_i$ covers a target $t_j$ if the $t_j$ lies within the sensing range of $s_i$. A sensor cover S is a set of sensors that jointly cover all the targets. Formally, $S = \{s_i \mid$ for each $t_j$ there is a $s_i \in S$ such that $s_i$ covers $t_j\}$. The life time of a sensor cover S, x(S), cannot exceed $\text{Min}_{s_i \in S} b_i$.

Energy-efficient target coverage problem [7] is to maximize sum of $x(S_k)$ (we shall refer to $x_k$) for all sensor covers with the constraints that no sensor $s_i$ can be used longer than $b_i$, the initial battery life. The problem is to find a family of sensor covers (with non-zero x(S)) which has maximum aggregated lifetime among all families of covers. In other words, the set of all targets are monitored by the network for the longest period of time with the given constraints of $b_i$ as the battery life of sensor $s_i$. The problem is shown to be NP-complete [7].

**Definition 1**

Define a matrix C as follows.

$$C_{ij} = \begin{cases} 1, & \text{if sensor } s_i \text{ is in sensor cover } S_j \\ 0, & \text{Otherwise} \end{cases}$$

The linear programming formulation of the energy efficient target coverage problem [7] can be stated as follows.

$$\text{Maximize } \sum_p x_p$$
$$\text{subject to } \sum_p C_{ip} x_p \leq b_i \quad \text{for all sensors } s_i$$
$$x_p \geq 0, \text{ for all sensor covers } S_p.$$

The constraint matrix C is explicitly known if the set of all sensor covers is known in advance. This is not practical. Thus the conventional algorithm for solving the linear programming cannot be employed as C is not known. Moreover, the number of columns of C is prohibitory large. The other approach is to generate covers (columns of C) as and when necessary (the typical column generation method of linear programming).

## 3. EARLIER RESULTS

The problem of coverage in wireless sensor networks has been studied from many different aspects: area coverage and point coverage. In area coverage problem, the given area is said to

International Journal of Ad hoc, Sensor & Ubiquitous Computing (IJASUC) Vol.2, No.1, March 2011

be covered if each point of the area is monitored by at least one sensor. The research work in [2, 3, 5, 19] consider a large number of sensors, deployed randomly for area monitoring. Cardei et al [2] model the disjoint sets as disjoint dominating sets in an undirected graph and a graph-coloring mechanism is proposed for computing the disjoint dominating sets. Slijepcevic and Potkonjak [3] model the area as a collection of fields, where every field has the property that any enclosed point is covered by the same set of sensors. The most-constrained least-constraining algorithm [3] computes the disjoint covers successively by selecting sensors that cover the critical element (field covered by a minimal number of sensors), giving priority to sensors that cover a high number of uncovered fields, cover sparsely covered fields and do not cover fields redundantly. Berman et al [5] propose a $(1+\epsilon)f$-approximation based on Garg and Konemann approximate scheme.

**Energy-Efficient Target Coverage**

In this paper, we survey recent contributions addressing energy-efficient coverage problems in the context of static WSNs, in which sensor nodes do not move once they are deployed. Sensors can monitor a disk centered at the sensor's location, whose radius equals the sensing range. Target coverage problem presented in [7, 9, 20, 24, 25, 29] attempts to maximize the total network lifetime by grouping sensors into non-disjoint sensor covers and then activating them one after another. Cardei et al [7] propose a greedy heuristic to solve target coverage problem where heuristic gives priority to that sensor which covers critical target (target covered by least number of sensors) and maximum number of uncovered targets while generating sensor covers. In [9], authors maximize the lifetime for target coverage problem by organizing the sensors into maximum disjoint set covers and these are activated successively. Recently, Zorbas et al [20] propose another algorithm for target coverage problem where all deployed sensors are grouped into four sub classes: *Best, Good, Ok* and *Poor* based on the coverage quality. The proposed approach always tries to select sensors from the *Best* class first and if there are no more sensors in best class then it selects from *Good* and so on. In [25] authors proposed a column generation based heuristic solution for target coverage by finding more patterns randomly to get initial basic feasible set (BFS). Gu Y et al [29] proposed a pattern based solution for target coverage problem to prolong the total network lifetime.

**Energy-Efficient Connected Coverage**

An important issue in WSN is connectivity. A network is connected if any active node can communicate with any other active node, possibly using intermediate nodes as relays. Once the sensors are deployed, they organize into a network that must be connected so that the information collected by sensor nodes can be relayed back to data sinks or controllers. An important, frequently addressed objective is to determine a minimal number of working sensors required to maintain the initial coverage area as well as connectivity. Selecting a minimal set of working nodes reduces power consumption and prolongs network lifetime. Next we will present several connected coverage mechanisms.

The research work presented in [12, 21, 23, 26, 27, 28] discuss a variant of target coverage problem known as target connected coverage where all the sensors in generated sensor cover must be connected to some designated base station (BS) with the help of some relay nodes. Jaggi et al [12] proposed a connected cover set generation algorithm in order to extend the lifetime of the network with the consideration that all the cover sets are disjoint and they try to maximize their number, while computing a shortest path tree to select the relay nodes that manage to retain connectivity in the network. Cardei et al [21] proposed centralized and distributed algorithms for the computation of the connected cover sets. A breadth first search algorithm is used to discover the node-path to the BS through a centralized algorithm, while a minimum spanning tree algorithm is used in the distributed version of the algorithm. Lu et al [23] presents an approach for target coverage and connectivity by presenting a distributed algorithm that builds a virtual backbone first to satisfy network connectivity, and it ensure coverage based on that backbone. D. Zorbas et al [27] presented another heuristic solution for target coverage. They proposed static as well dynamic heuristic for the given coverage problem. S Begum et al [28] also proposed a heuristic solution based on Ant Colony Problem for target connected coverage in wireless sensor networks.

### Target Coverage Under QoS Constrain

Some applications may require different degrees of coverage while still maintaining working node connectivity. We say that a network has a coverage degree k (k-coverage) if every target is within the sensing range of at least k-sensors. Networks with a higher coverage degree can obtain higher sensing accuracy and be more robust to sensor failure.
The K-coverage problem is introduced by Gu et al [18, 22] where the domatic partition problem is discussed in which every target constantly covered by *K*- number of sensors.

### Target Coverage with Adjustable Sensing Ranges

There is one more variant of target coverage problem that is based on multiple sensing ranges [8, 11] of deployed sensors. Each sensor in WSN has *P* sensing ranges and based on the coverage requirement, at a time one sensing range is chosen by sensor. M. Cardei et al [8] have studied the target coverage problem with multiple sensing ranges, i.e., how to schedule the sensors and their sensing ranges to maximize the lifetime of network providing that all the targets are covered by sensors. M. Cardei [8] *et al:* firstly prove that the problem is NP-Complete. In order to solve the problem, a centralized greedy and heuristic algorithm named CGH and a distributed localized greedy and heuristic algorithm named DLGH are proposed in [8]. Sensors with large contributions are always selected with a high priority in CGH. This may result in phenomena: some sensors will die out too early because of their large contributions. Only local information can be used in distributed DLGH, which may result in many redundant sensors in each round too. A. Dhawan et al [11] provided an alternative problem formulation for the lifetime maximization problem in a sensor network with adjustable sensing ranges. They present a packing LP formulation of the problem and give a heuristic solution (both centralized and distributed).

### Centralized and Distributed Approach

There are two different types of algorithms: distributed and centralized. In centralized, algorithms are always executed at a powerful center such as Base Station (BS), after that the result is scattered to each sensor in the network. Centralized algorithms in [3, 5, 7, 20, 21] formulate the problem as the maximization of the network lifetime under the area [3, 5] and target [7, 20, 21] coverage constraint. In distributed algorithms [4, 9, 12, 21, 23], the decision process is decentralized where a number of sensor nodes perform the required task cooperatively and then they disseminate the scheduling information to the rest of the sensors. In this paper we restrict our scope to the centralized algorithm.

### Disjoint and non Disjoint Approach

One method for extending the sensor network lifetime through energy resource preservation is the division of the set of sensors into disjoint sets where each sensor can participate only in one set cover. These disjoint sets are activated successively, such that at any moment in time only one set is active. As all targets are monitored by every sensor set, the goal of the approaches [2, 3, 27] is to determine a maximum number of disjoint sets, so that the time interval between two activations for any given sensor is longer.

In the case of non-disjoint algorithms, nodes may participate in more than one cover sets. In some cases, this may prolong the lifetime of the network in comparison to the disjoint cover set algorithms. Moreover, non-disjoint algorithms [5, 7, 9, 20] may generate more cover sets than node-disjoint ones, but the generating algorithm incurs a higher order of complexity.
There are two known heuristics [5], [7] that use column generation technique to solve target coverage problem. It is interesting to note that though both the algorithms look different, they adopt the same algorithmic principle. From the original proposals of the algorithms, it is not apparent that these two algorithms share some commonalities.

## 4. MOTIVATION

The LP formulation of target coverage problem is presented by M. Cardei et al [7]. As solving the LP problem is computationally complex because of unpredictable size of matrix C. All the approaches discussed in literature try to maximize the total network lifetime. So, in order to maximize the total network lifetime, we propose to design an energy-efficient heuristic solution for the given target coverage problem. For the sake of explaining our heuristic, we dedicate this section introduce some preliminary knowledge, definitions and to state our assumptions.

**Definition 2 (sensor-target coverage matrix)**

Define n×m sensor-target coverage matrix M as follows.

$$M_{ij} = \begin{cases} 1, & \text{if sensor } s_i \text{ covers target } t_j \\ 0, & \text{otherwise} \end{cases}$$

**Definition 3 (sensor cover)**

Given M, a sensor cover $S$ is a set of rows of M such that for every column $j$, there is a row $i$ in $S$ such that $M_{ij} = 1$. A sensor cover $S$ is a *minimal cover* if for any cover S', $S' \subseteq S$ if and only if $S' = S$.

**Definition 4 (maximum allowable lifetime of sensor cover $S$)**

The maximum allowable lifetime of a sensor cover $S$ is the smallest available lifetime of its sensors. Thus
$$max\_lifetime(S) = \underset{si \in S}{Min} \, b_i$$

**Definition 5 (critical target)**

A target $t_j$ is said to be critical target if $j = \arg\min \Sigma_i M_{ij} b_i$

**Definition 6 (Upper bound)**

Let us define a quantity $u$ as follows: $u = \underset{j}{Min} \sum_i M_{ij} b_i$

To build our arguments, we first describe a generic greedy algorithm that generates a sensor cover and assigns maximum allowable lifetime to covers. Henceforth we assume the initial value of $b_i = 1$ and at any stage of the algorithm, $b_i$ is the residual life time of the sensor $s_i$. *Algorithm Naïve Greedy Algorithm* gives the pseudo code of the simple greedy heuristic. The value of $w$ is between 0 and 1.

```
INITIALIZATION   C = ∅;   S_cur = all sensors
   while  S_cur ≠ ∅  do
      generate a cover
         greedy_cover_generate
            //generates a sensor cover S in some greedy fashion and returns NULL if no cover is found //
         if S ≠ NULL
            C ← C ∪ {S}
            lifetime assignment
            x(S) ← max_lifetime(S)
            update
            for all  s_i ∈ S   update  b_i ← b_i – w
               if b_i = 0 then  S_cur ← S_cur − {s_i}
         else, return
   end do
```

Naïve Greedy Algorithm

# 5. HIGH-ENERGY-FIRST (HEF) HEURISTIC

In this section, based on above observations we propose a new heuristic to solve the target coverage problem. We observe that the granularity parameter $w$ plays an important role in getting a better approximation of optimal solution. Hence prioritizing the sensors in terms of residual battery provides us better opportunity of using the sensors. HEF uses the three important steps in the following manner.

### Generate a cover

The HEF heuristic generates sensor cover $S$ by selecting a sensor with highest residual battery life and which covers at least one uncovered target. Ideally, some sort of priority (weight) is associated with each sensor. The sensor cover is constructed by iteratively selecting sensors of high priorities till all the targets are covered.

### Assign lifetime to a cover

For a sensor cover $S$ generated in the previous step, we assign some lifetime x($S$). As shown in naïve greedy, the temptation is to assign maximum allowable lifetime. Instead, the algorithm requires a user-specified constant $w$ and whenever a cover S is generated, $w´ = Min(w, max\_lifetime(S))$ is added to its lifetime. By this process, we do not consume the total energy of sensors and make these sensors available for other covers.

### Change the priorities of the sensors

In order to avoid the repeated generation of the same sensor cover in consecutive iterations, the priority of a sensor reduces once it is used in a sensor cover and as a result the greedy construction of sensor cover in the next iteration tries to avoid such a sensor.

```
INITIALIZATION   C = ∅;   S_cur = all sensors
   while  S_cur ≠ ∅  do
      generate a cover HEF
            //generates a sensor cover S  and returns NULL if no cover is found //
         initialise S = ∅
         T_uncovered = T
         do while  T_uncovered ≠ ∅
            select a sensor s with maximum b_i (= residual battery life) among sensors that cover at least one uncovered target
               S ← S ∪ {s}
               for all target t covered by s,
```

```
                T_uncovered = T_uncovered -{t}
        end do
        minimalize S
          if S ≠ NULL
             C ← C ∪ {S}
             lifetime assignment
             x(S) ← max_lifetime(S)
             update
             for all s_i∈ S  update  b_i ← b_i – x(S)
                if b_i = 0 then  S_cur ← S_cur − {s_i}
                   else, return
     end do
     end while
```

*generate a cover HEF*

Sensor cover generation step in HEF is different from all other three discussed algorithms, namely Greedy Heuristics by Cardei et al [7], Garg-Konemann based approximation scheme by Berman et al [5], and BGOP by Zorbas et al [20].

We do not use the concept of critical targets as [5, 7, 20] because this is not possible without price. Our proposed method generates nonminimal sensor covers and hence we need an additional step to minimalize the cover. We are minimalizing the cover by removing one sensor at a time from obtained cover and then checking whether it is a cover or not, if still it is a cover then we can remove this sensor from the current set cover. After repeating this process for all sensors which are in the current sensor cover, we get the minimal cover. We give pseudocode of algorithm *generate a cover HEF*. In the following section we show that HEF gives better solution for randomly generated problem instances.

## 6. EXPERIMENTAL ANALYSIS

For the experimental study, we implemented the heuristic in c programming language. All experiments were carried out on a Pentium (4) 2.63GHz host with 256 MB of RAM, running Window Xp/Linux operating system. We generate the problem instances randomly. We assume a sensing area of 800×800m inside the monitored area of 1000×1000m. We assume that sensors are homogenous and initially have the same energy and have similar sensing range. Sensors and targets are generated in terms of their coordinates which are generated by Pseudo-random number generation routine. We assume that if the Euclidean distance of the target from a sensor is less than 70m, then the target falls within the sensing range of the sensor. For our experiments, we vary the number of sensors in interval [20, 150], number of targets in [20, 90]. The upper bound on network lifetime can be calculated as definition 6 in section 4. During Experimentation, **Algorithm 1**: Greedy Heuristic [7], **Algorithm 2**: Approximation Scheme [5] and **Algorithm 3**: BGOP [20] will be assumed.

### Experiment 1

This experiment tries to answer "How good is Naïve greedy technique?" Naïve greedy works for $w$ =1 (maximum allowable lifetime) which makes it different from generic greedy. Algorithm 1 and HEF are experimented for different values of $w$ to find the heuristic solution closer to the optimal solution. The experiment clearly shows that total network lifetime can be significantly improved when we permit values of w which can be less than 1. But, even experimenting with smaller $w$ (0.002), the difference between the lifetime obtained by smaller w =.0.002 and with $w$ =1 is not that much high and due to this $w$ =1 is quite acceptable.

In this experiment we have taken six different values of $w$ as shown in Table 1 and Table 2. We experimented for fix (150) sensor nodes and varying target points between 20 and 90 with an

51

International Journal of Ad hoc, Sensor & Ubiquitous Computing (IJASUC) Vol.2, No.1, March 2011

increment of 10. The average of lifetime obtained by Algorithm 1 for 15 random problem instances is presented by Table 1. Table 2 shows the same experiment for HEF algorithm.

Table 1: The average lifetime computed by Algorithm 1 for different w

| Target | $w=1$ | $w=0.50$ | $w=0.25$ | $w=0.025$ | $w=0.01$ | $w=0.002$ | A | B | C |
|---|---|---|---|---|---|---|---|---|---|
| 20 | 65.1 | 65.25 | 65.45 | 65.36 | 65.429 | 65.437 | 65.1 | 65.45 | 0.54 |
| 30 | 57.8 | 58.5 | 58.7 | 58.79 | 58.806 | 58.7166 | 57.8 | 58.806 | 1.72 |
| 40 | 56.6 | 56.95 | 56.525 | 56.6825 | 56.641 | 56.645 | 56.525 | 56.95 | 0.75 |
| 50 | 56.3 | 56 | 56.2 | 55.905 | 55.903 | 55.9018 | 55.9018 | 56.3 | 0.71 |
| 60 | 56.4 | 55.85 | 55.625 | 55.3925 | 55.413 | 55.4004 | 55.3925 | 56.4 | 1.81 |
| 70 | 56.6 | 56.75 | 56.675 | 56.5645 | 56.559 | 56.5524 | 56.5524 | 56.75 | 0.35 |
| 80 | 57.7 | 57.8 | 57.875 | 57.5975 | 57.571 | 57.5856 | 57.571 | 57.875 | 0.53 |
| 90 | 56.4 | 56.45 | 56.625 | 56.4525 | 56.41 | 56.4093 | 56.4 | 56.625 | 0.40 |

A = Minimum lifetime obtained, B = Maximum lifetime obtained, C = (B-A)/ (Average lifetime) %

Table 2: The average lifetime computed by HEF for different *w*

| Target | $w=1$ | $w=0.50$ | $w=0.25$ | $w=0.025$ | $w=0.01$ | $w=0.002$ | A | B | C |
|---|---|---|---|---|---|---|---|---|---|
| 20 | 68.2 | 69.2 | 69.65 | 70.1375 | 70.1745 | 70.1948 | 68.2 | 70.1948 | 2.86 |
| 30 | 61.2 | 62.2 | 62.625 | 63.0925 | 63.127 | 63.1374 | 61.2 | 63.1374 | 3.10 |
| 40 | 59.7 | 60.5 | 60.725 | 60.9245 | 60.94 | 60.948 | 59.7 | 60.948 | 2.05 |
| 50 | 59.4 | 59.9 | 60.125 | 60.35 | 60.39 | 60.304 | 59.4 | 60.39 | 1.65 |
| 60 | 59 | 59.65 | 59.9 | 60.06 | 60.054 | 60.0518 | 59 | 60.06 | 1.77 |
| 70 | 59 | 59.5 | 59.675 | 59.84 | 59.87 | 59.8952 | 59 | 59.8925 | 1.50 |
| 80 | 59.3 | 60.1 | 60.375 | 60.63 | 60.642 | 60.6484 | 59.3 | 60.6484 | 2.24 |
| 90 | 57.6 | 58.4 | 58.9 | 59.3975 | 59.428 | 59.4458 | 57.6 | 59.4458 | 3.14 |

A = Minimum lifetime obtained, B = Maximum lifetime obtained, C = (B-A)/ (Average lifetime) %

In table 1, we can observe that for Algorithm 1 the maximum increment in network lifetime obtained with smallest *w* =0.002 is 1.81% when compared with average lifetime obtained with highest *w* =1. The similar observation comes for HEF (Table 2) where the maximum increment is 3.14%.Hence, because of little improvement in total lifetime obtained by Algorithm 1 and HEF using smaller *w*, we can say that the lifetime obtained with *w* =1(Naïve greedy) is also acceptable.

increment of 10. The average of lifetime obtained by Algorithm 1 for 15 random problem instances is presented by Table 1. Table 2 shows the same experiment for HEF algorithm.

Table 1: The average lifetime computed by Algorithm 1 for different w

| Target | $w=1$ | $w=0.50$ | $w=0.25$ | $w=0.025$ | $w=0.01$ | $w=0.002$ | A | B | C |
|---|---|---|---|---|---|---|---|---|---|
| 20 | 65.1 | 65.25 | 65.45 | 65.36 | 65.429 | 65.437 | 65.1 | 65.45 | 0.54 |
| 30 | 57.8 | 58.5 | 58.7 | 58.79 | 58.806 | 58.7166 | 57.8 | 58.806 | 1.72 |
| 40 | 56.6 | 56.95 | 56.525 | 56.6825 | 56.641 | 56.645 | 56.525 | 56.95 | 0.75 |
| 50 | 56.3 | 56 | 56.2 | 55.905 | 55.903 | 55.9018 | 55.9018 | 56.3 | 0.71 |
| 60 | 56.4 | 55.85 | 55.625 | 55.3925 | 55.413 | 55.4004 | 55.3925 | 56.4 | 1.81 |
| 70 | 56.6 | 56.75 | 56.675 | 56.5645 | 56.559 | 56.5524 | 56.5524 | 56.75 | 0.35 |
| 80 | 57.7 | 57.8 | 57.875 | 57.5975 | 57.571 | 57.5856 | 57.571 | 57.875 | 0.53 |
| 90 | 56.4 | 56.45 | 56.625 | 56.4525 | 56.41 | 56.4093 | 56.4 | 56.625 | 0.40 |

A = Minimum lifetime obtained, B = Maximum lifetime obtained, C = (B-A)/ (Average lifetime) %

Table 2: The average lifetime computed by HEF for different *w*

| Target | $w=1$ | $w=0.50$ | $w=0.25$ | $w=0.025$ | $w=0.01$ | $w=0.002$ | A | B | C |
|---|---|---|---|---|---|---|---|---|---|
| 20 | 68.2 | 69.2 | 69.65 | 70.1375 | 70.1745 | 70.1948 | 68.2 | 70.1948 | 2.86 |
| 30 | 61.2 | 62.2 | 62.625 | 63.0925 | 63.127 | 63.1374 | 61.2 | 63.1374 | 3.10 |
| 40 | 59.7 | 60.5 | 60.725 | 60.9245 | 60.94 | 60.948 | 59.7 | 60.948 | 2.05 |
| 50 | 59.4 | 59.9 | 60.125 | 60.35 | 60.39 | 60.304 | 59.4 | 60.39 | 1.65 |
| 60 | 59 | 59.65 | 59.9 | 60.06 | 60.054 | 60.0518 | 59 | 60.06 | 1.77 |
| 70 | 59 | 59.5 | 59.675 | 59.84 | 59.87 | 59.8952 | 59 | 59.8925 | 1.50 |
| 80 | 59.3 | 60.1 | 60.375 | 60.63 | 60.642 | 60.6484 | 59.3 | 60.6484 | 2.24 |
| 90 | 57.6 | 58.4 | 58.9 | 59.3975 | 59.428 | 59.4458 | 57.6 | 59.4458 | 3.14 |

A = Minimum lifetime obtained, B = Maximum lifetime obtained, C = (B-A)/ (Average lifetime) %

In table 1, we can observe that for Algorithm 1 the maximum increment in network lifetime obtained with smallest *w* =0.002 is 1.81% when compared with average lifetime obtained with highest *w* =1. The similar observation comes for HEF (Table 2) where the maximum increment is 3.14%.Hence, because of little improvement in total lifetime obtained by Algorithm 1 and HEF using smaller *w*, we can say that the lifetime obtained with *w* =1(Naïve greedy) is also acceptable.

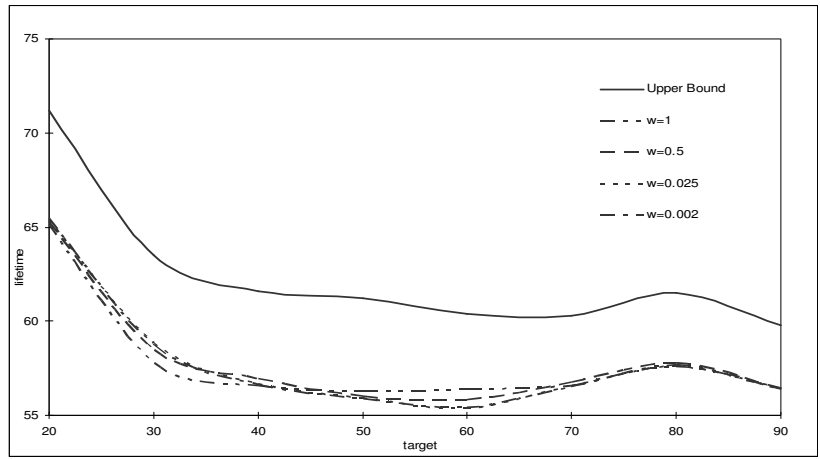

Figure 1: Average lifetime computed by Algorithm 1 for different values of *w* and the upper bound on the optimal solution

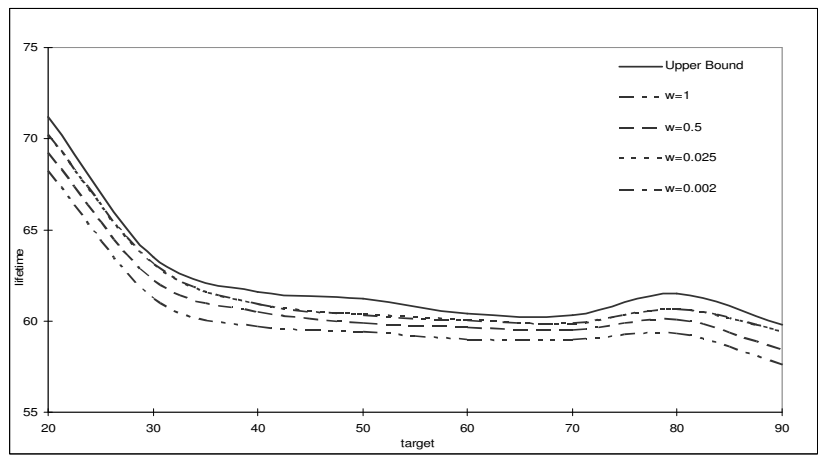

Figure 2: Average lifetime computed by HEF for different values of *w* and the upper bound on the optimal solution

For the given problem instances the optimal value is not known but the upper bound (Definition 6) of the optimal can be calculated. The actual optimal is obtained when heuristic value reaches the upper bound. The above graphs (Figure 1 and 2) summarize Experiment 1. For the clarity purpose graph are drawn only for four *w* among six *w*. Figure 1 gives the average lifetime obtained by Algorithm 1 with respect to upper bound and with respect to *w*. Figure 2 gives the same analysis for HEF.

We observe that for Algorithm 1, there is no particular advantage of varying *w*. Moreover, for *w* =1, the algorithm gives acceptable solution (within 10% of the upper bound). Like Algorithm 1, for HEF there is no advantage of fine granularity, but it is interesting to note that for all values of *w*, HEF gives solution very close to the upper bound. In fact in majority of the cases we get the exact optimal solution. Our main inferences in Experiment 1 are- (1) Naïve Greedy can as well be used to get acceptably-close-to-optimal solution and (2) HEF yields better solution than Algorithm 1.

**Experiment 2**

In this experiment we try to compare Algorithm 2 with other algorithms, namely Algorithm 1 and HEF. Though Algorithm 2, as originally proposed, seems to be based on a different paradigm, we have justified here that this algorithm also follows the generic greedy heuristic. The following relation defines correspondence between $w$ of other algorithms and $\varepsilon$ of Algorithm 2.

$$w = \frac{\varepsilon \ln(1+\varepsilon)}{\ln((1+\varepsilon)n)}$$

In order to compare the performance of Algorithm 2 with other algorithms, experiments were carried out for different values $w$ (and equivalent values of $\varepsilon$) for different set of problem instances. We experimented for fix (150) sensor nodes and varying targets between 20 and 90 with an increment of 10. The graphs in Figure 3 and 4 summarize the results for two specific values of $w$, namely Figure 3- for $\varepsilon=0.104$ and $w=0.002$ and Figure 4- for $\varepsilon=0.25$ and $w=0.01$.

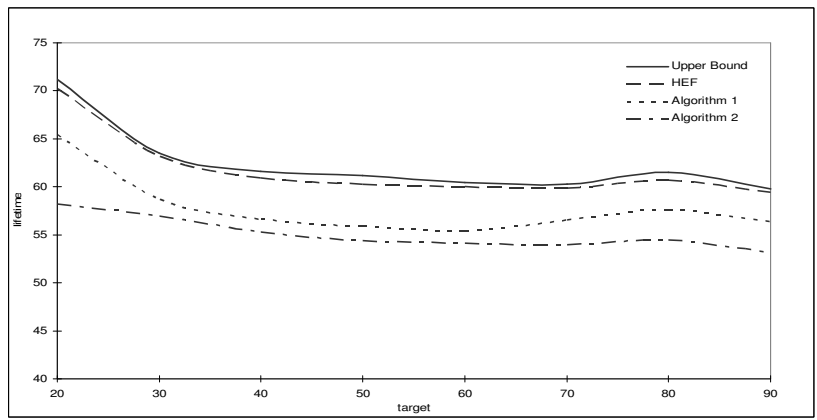

Figure 3: Average solutions obtained by Algorithm1, Algorithm 2, HEF and the upper bound and for $\varepsilon=0.104$ and $w=0.002$

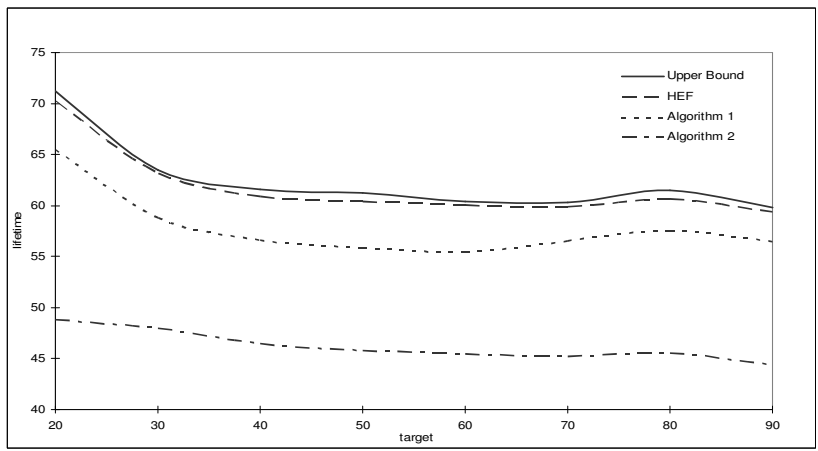

Figure 4: Average solutions obtained by Algorithm1, Algorithm 2, HEF and the upper bound and for $\varepsilon=0.25$ and $w=0.01$

**Experiment 3**

In this experiment we investigate the overall performance of HEF with respect to Algorithm 1 and Algorithm 3. We did not consider Algorithm 2 in this experiment as for higher values of *w*, the corresponding ε are too high for Algorithm 2 to proceed beyond the first iteration. On the other hand, very low value of *w* is not practical for Algorithm 1 or Algorithm 3.

Figure 5 gives the average of lifetime obtained by HEF, Algorithm 1 and Algorithm 3 for 15 random problem instances for different number of targets. The number of sensors is fixed to 150 and we experimented for *w* equals 1.

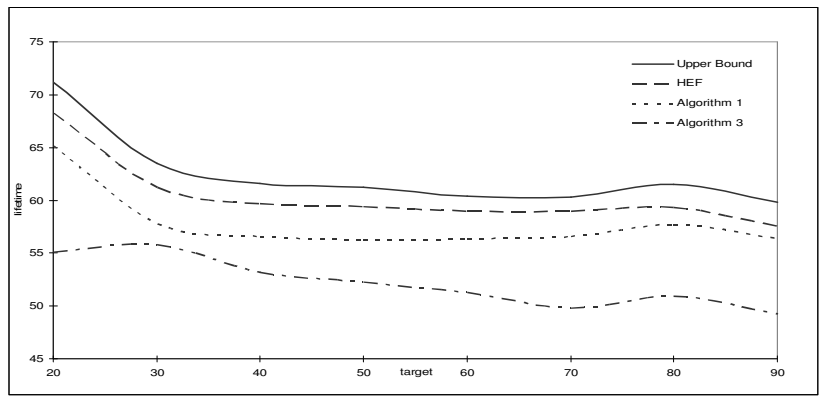

Figure 5: Average lifetime obtained by HEF, algorithm 1 and Algorithm 3 with *w* =1

We can observe from Figure 5 that the average network lifetime obtained by HEF is very close to the average upper bound on lifetime and the performance of HEF is almost like a replica of optimal value.

**Experiment 4**

In experiment 4, we carry out experiment of the similar nature while varying the number of sensors and fixing the number of targets to 25. Figure 6 shows the average lifetime obtained by different algorithms for 15 random problem instances against the upper bound.

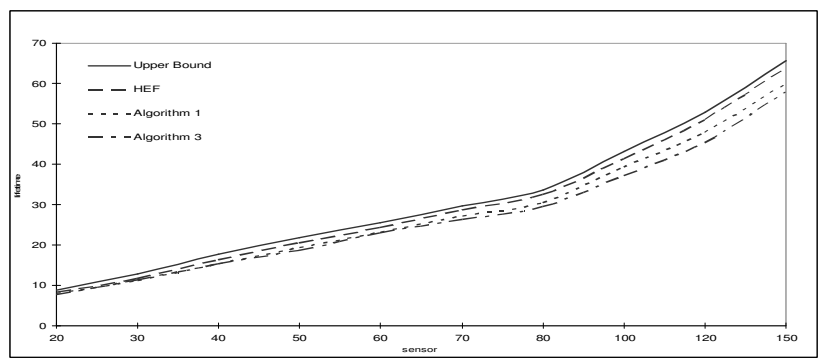

Figure 6: Average lifetime by Algorithm 1, Algorithm 3 and HEF and the upper bound on the optimal solution.

We observe from figure 6 that the HEF always outperforms Algorithm 1 and Algorithm 3.

When we compare experiment 4 with experiment 1, experiment 2 and experiment 3, then we can observe that in wireless sensor networks the total network lifetime increses with sensor node density because this time more sensors are covering same target points which results in more set coers and obtain longer network lifetime. Likewise, As the number of targets grows, the average number of sensors that cover every target decreses, resulting in fewer covers. So, to carefully monitor all target points we have to deploy sufficient number of sensor nodes which results in higher total network lifetime. Opting small *w* is being prefered over higher *w* because for small *w*, a particular sensor can participate in more set covers and remains active for longer time which results in higher network lifetime and we have already observed from Figure 2 that HEF provides near optimal lifetime for smaller *w*.

## 7. CONCLUSIONS

In this paper we study centralized algorithms for the energy efficient target coverage problem. We show that three major algorithms are based on a common algorithmic principle. Though one of the algorithms makes some claim on the quality of solution, it is not a practical bound for large number of sensors. We show that a simple, naïve technique can be utilized to get a reasonably good solution.  We propose a new heuristic, High-Energy-First (HEF) which prioritizes sensors based on their residual battery life. We show empirically that our algorithm provides better solution than other algorithms. We present different aspects of experimental analysis to establish this observation.

In this paper the basic target coverage problem is considered. There have been many variants of this problem such as target coverage with adjustable sensing range or oriented sensing etc [6, 10, 11, 15, 17]. We propose to extend our study to different variants.  Maximizing the total lifetime of the network in a distributed setting is yet another direction of future research. Another important problem is scheduling the individual sensors' active/sleep duration based on the output of target coverage problem [3, 12, 16].

### ACKNOWLEDGEMENTS

The authors thank Prof S K Gupta for useful discussions. The first author also thanks Subrat Dash for providing useful hints during experimentation.

International Journal of Ad hoc, Sensor & Ubiquitous Computing (IJASUC) Vol.2, No.1, March 2011

**Authors**

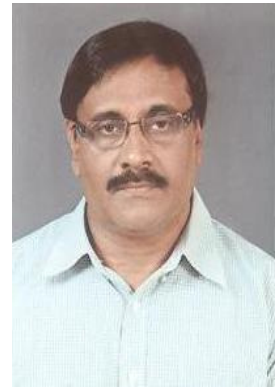

Arun K Pujari is Professor of Computer Science at the University of Hyderabad, Hyderabad. Prior to joining UoH, he served at the Automated Cartography Cell, Survey of India, and Jawaharlal Nehru University, New Delhi. He received his PhD from the Indian Institute of Technology, Kanpur and MSc from Sambalpur University, Sambalpur. He has also undertaken several visiting assignments at the Institute of Industrial Sciences, University of Tokyo; International Institute of Software Technology, United Nations University, Macau; University of Memphis, USA; and Griffith University, Australia, among others. Professor Pujari is at present the vice-chancellor of Sambalpur University.
His research area is AI, GIS, Combinatorial Algorithms, Data Mining.

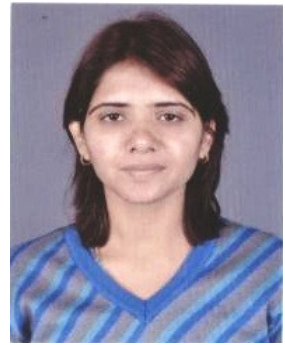

Manju has received her Master's degree and B. E. in Computer Science in 2009 and 2006 respectively. Recently, she is working as Lecturer at SUIIT, Sambalpur, Odisha. Prior to join this, she was Lecturer at BITS Pilani. Her research interest is Algorithm designing, Wireless Sensor Networks.
.